# How Criticality of Gene Regulatory Networks Affects the Resulting Morphogenesis under Genetic Perturbations

Hyobin Kim and Hiroki Sayama

Department of Systems Science and Industrial Engineering

Center for Collective Dynamics of Complex Systems

Binghamton University, State University of New York, Binghamton, NY, 13902 USA

hkim240@binghamton.edu

**Abstract** Whereas the relationship between criticality of gene regulatory networks (GRNs) and dynamics of GRNs at a single cell level has been vigorously studied, the relationship between the criticality of GRNs and system properties at a higher level has not fully explored. Here we aim at revealing a potential role of criticality of GRNs at a multicellular level which are hard to uncover through the single-cell-level studies, especially from an evolutionary viewpoint. Our model simulated the growth of a cell population from a single seed cell. All the cells were assumed to have identical GRNs. We induced genetic perturbations to the GRN of the seed cell by adding, deleting, or switching a regulatory link between a pair of genes. From numerical simulations, we found that the criticality of GRNs facilitated the formation of nontrivial morphologies when the GRNs were critical in the presence of the evolutionary perturbations. Moreover, the criticality of GRNs produced topologically homogenous cell clusters by adjusting the spatial arrangements of cells, which led to the formation of nontrivial morphogenetic patterns. Our findings corresponded to an epigenetic viewpoint that heterogeneous and complex features emerge from homogeneous and less complex components through the interactions among them. Thus, our results imply that highly structured tissues or organs in morphogenesis of multicellular organisms might stem from the criticality of GRNs.

**Keywords**

# 1. Introduction

Random Boolean networks (RBNs) have been used extensively as a theoretical model of gene regulatory networks (GRNs) in artificial life and complex systems research [1, 3, 13, 14, 25, 30, 38, 43]. In the context of GRNs, the concept of *criticality* has been discussed as a phase transition point between ordered and chaotic regimes for the dynamics of those networks [19, 20]. The criticality of GRNs has been recognized as a property which makes robustness and adaptability coexist in living organisms [2]. When perturbations are added to GRNs, ordered GRNs are so robust that they just sustain existing cellular functions. On the contrary, chaotic GRNs are so adaptable that they severely create new functions rather than conserving existing ones. Meanwhile, critical GRNs stably sustain their functions against the perturbations, and at the same time flexibly generate new phenotypes which help organism to adapt to new environments because they have optimal balance between robustness and adaptability.

Whereas many studies for elucidating the relationship between criticality of GRNs and dynamics of GRNs at a single cell level have been performed [4, 29, 32, 36, 37, 39], the relationship between the criticality and properties at a higher level has not fully explored. Only a few studies have figured out how the properties of GRNs influence properties at a multicellular level [6, 7, 12, 26, 41]. They all used a discrete space like Cellular Automata, which is not realistic as a model of morphogenetic processes. To investigate a potential role of criticality of GRNs at a multicellular level which is hard to uncover through the single-cell-level studies, we have recently proposed GRN-based morphogenetic systems operating in a continuous space and revealed that the criticality of GRNs facilitated the formation of nontrivial morphologies [21, 22]. To make our finding more solid from an evolutionary viewpoint, in this study, we include an evolutionary process in our morphogenetic model by introducing genetic perturbations (e.g., mutations) to GRNs. We add perturbations to GRNs by adding, deleting, or switching a regulatory link between a pair of genes. We look into whether the role of the criticality of GRNs can be maintained even in the presence of the evolutionary perturbations and what the resulting morphologies are like.

The rest of the paper is structured as follows. In Section 2, we provide a brief literature review concerning the relationship between criticality of GRNs and dynamics of GRNs. In Section 3, we describe our GRN-based morphogenetic model. In Section 4, parameter setting for simulations and various measures for results analysis are explained. In Section 5, morphologies obtained from the simulations of the morphogenetic model are analyzed. Section 6 summarizes and concludes the paper.

## 2. Literature Review

Since the notion of criticality of GRNs based on RBNs was established by Kauffman [19, 20], numerous studies have been conducted on whether GRNs of living organisms are dynamically critical or not. The studies introduced in this section compare dynamic behaviors of RBNs in the critical regime with gene expression dynamics (Serra et al. [36], [37], Rämö et al. [32], Shmulevich et al. [39], and Nykter et al. [29]) or dynamics of Boolean models of genetic networks (Balleza et al. [4]), both of which are based on gene expression data of real living organisms. They demonstrate that the dynamics of the living organisms are consistent with those of critical RBNs, and thus conclude that the dynamics of living organisms are critical.

Serra et al. [36] showed that the gene expression dynamics of *Saccharomyces cerevisiae* (commonly known as baker's yeast) were critical through the comparison of critical RBNs and the gene expression data of *Saccharomyces cerevisiae* in the perturbation avalanches analysis, which measured the size of an avalanche, i.e., the number of genes that are affected by the knockout of a single gene. They found that the distribution of avalanche sizes of critical RBNs approximated to that of *Saccharomyces cerevisiae*. Also, Serra et al. [37] demonstrated the numerical result analytically. Similarly, Rämö et al. [32] displayed that the gene expression dynamics of *Saccharomyces cerevisiae* are critical using approximate formulas for the distributions of avalanche sizes. They showed that the distributions of avalanche sizes of both critical RBNs and *Saccharomyces cerevisiae* followed power-law distributions with the same exponent value.

Shmulevich et al. [39] and Nykter et al. [29] showed that the gene expression dynamics of biological systems exhibited criticality applying quantitative measures used in data compression to the gene expression data of living organisms. Shmulevich et al. [39] measured Lempel-Ziv (LZ) complexity of both the binarized gene expression data during Hela cell cycle progression and RBNs in ordered, critical, and chaotic regimes. They found that the LZ value obtained from the gene expression dynamics of Hela Cells was consistent with that of either ordered or critical dynamic behavior. Nykter et al. [29] calculated normalized compression distance (NCD) from the gene expression data of macrophage. They compared the measured values with the NCD values of ordered, critical, and chaotic RBNs. They found that the trajectory of NCD of macrophage corresponded with that of critical RBNs.

Balleza et al. [4] showed that the dynamics of living organisms in four kingdoms operated close to criticality by examining the dynamics of Boolean models of examples belonging to four kingdoms in biology. Inferring interactions among genes using the gene expression data of *Saccharomyces cerevisiae* (yeast in kingdom fungi), *Escherichia coli* (bacteria in kingdom protista), *Bacillus subtilis* (bacteria in kingdom protista), *Drosophila melanogaster* (insect in kingdom animalia), and *Arabidopsis thaliana* (plant in kingdom plantae), they implemented five Boolean networks. They displayed that the slopes of Derrida curves of the five Boolean networks were similar to that of a critical RBN, where a Derrida plot visualizes the dynamic behaviors of Boolean networks [9, 10].

All the existing studies concentrate on the relationship between the criticality of GRNs and dynamics of GRNs at a single cell. In this study, we focus on revealing a potential role of criticality of GRNs in morphogenesis at a multicellular level under genetic perturbations.

## 3. Model

We have developed a computational model for morphogenetic processes of cell aggregation, in which all the cells have an identical GRN. Figure 1 shows the algorithm for our model. The simulation starts with one seed cell. In the cell, a RBN is produced as a GRN. To find neighboring cells, whether or not there are cells within the fixed neighborhood radius is examined. Through the interaction with neighbors, cells' fates are determined. We assume that there are four fundamental cell fates: proliferation, differentiation, apoptosis, and quiescence in our model. Cells expressing proliferation, differentiation, or quiescence can switch their fates through cell-cell interaction. The cells are positioned in a two-dimensional continuous space by spring-mass-damper (SMD) kinetics. Until the termination condition of the simulation is satisfied, one seed cell grows into an aggregation iterating the processes from "finding neighboring cells" to "positioning cells in the space" every time step. The simulator of our model was implemented in Java.

### 3.1 Gene Regulatory Network (GRN)

A RBN (a.k.a., $NK$ Boolean network) was suggested as a GRN model by Kauffman [18, 19, 20]. Here $N$ is the number of nodes and $K$ is the number of input links per node. One node means a gene. The state of a node can be either ON (1, activated) or OFF (0, inhibited). The node state is determined by the states of input nodes and Boolean function assigned to each node. A state space which is constructed from the topology of a RBN and assigned Boolean functions refers to the set of all the possible configurations. Figure 2 shows schematic diagrams for an example RBN and its state space. In the state space, stationary configurations are defined as attractors, and the others are called basins of attraction of the attractors. The dynamics of RBNs are divided into three regimes depending on the structure of state space: ordered, critical, and chaotic. Using node in-degree ($K$), internal homogeneity ($p$), or canalizing functions, the dynamics of RBNs can be varied. In the case of determining dynamics of a RBN using node in-degree ($K$), $K = 1$ is ordered, $K = 2$ is critical, and $K > 2$ is chaotic, on average [19, 20].

For our morphogenetic model, we use a RBN that consists of 16 nodes ($N = 16$) as a GRN. Adjusting node in-degree ($K$) from 1 to 4, we vary the dynamics of RBNs: ordered ($K = 1$), critical ($K = 2$), and chaotic ($K = 3, 4$). With in-vitro experimental data that attractors of GRNs represent cell types or cell fates, Huang suggested a conceptual framework to explain stochastic and reversible transitions between cell fates using $NK$ Boolean networks [5, 15, 16, 17]. Our morphogenetic systems are based on his framework. We randomly assign the four cell fates to attractors of GRNs. Concretely, if there is only one attractor, proliferation is assigned to the attractor. If there are two attractors, proliferation and differentiation are randomly assigned to the two attractors. Likewise, if there are three attractors, proliferation, differentiation, and apoptosis are randomly assigned to those attractors. If there are four or more attractors, proliferation, differentiation, and apoptosis are randomly assigned to three attractors and quiescence is assigned to the rest of the attractors (Figure 3).

For cellular behaviors, cells in proliferation are divided into two, where the daughter cells are placed within a fixed neighborhood radius ($R$) centering on the mother cells. Cells in differentiation are labeled as differentiated. Cells in apoptosis die and disappear. Cell in quiescence do not show any cellular behaviors.

### 3.2 Cell-Cell Interaction

Switching between cell fates occurs by perturbations of internal gene expression values of a GRN through cell-cell interaction. Our mechanism for cell-cell interaction is based on cell signaling of Damiani et al's multiple random Boolean networks on 2D cellular automata [6, 7]. In our model, a GRN has $n$ genes, which consist of $g$ normal genes and $r$ special genes as shown in Figure 4 (a). $r$ special genes exist in pairs where the genes producing signaling molecules ($r_1$) and receptors ($r_2$) are matched one to one. This one-to-one correspondence indicates signal transduction specificity by which certain signaling molecules respond to particular receptors. $r_1$ special genes synthesize signaling molecules and release them. Then, the corresponding $r_2$ receptors bind to the signaling molecules within the neighborhood radius ($R$).

The signal transduction has two mechanisms: autocrine and paracrine. Autocrine is a cell signaling where a cell produces signaling molecules and then receptors on the same cell bind to them. Paracrine is a cell signaling where a cell produces signaling molecules and then receptors on its neighboring cells bind to them. Figure 4 (b) explains the two mechanisms schematically. In our model, when there are no neighboring cells, autocrine is used. When there are neighbors, paracrine is used.

The gene expression values of a GRN are updated as follows:

- $g$ normal genes: the states of $g$ normal genes are updated by the states of input nodes and randomly assigned Boolean functions.
- $r_1$ genes producing signaling molecules: like $g$ normal genes, the states of $r_1$ genes are updated by the states of input nodes and randomly assigned Boolean functions. If the states of $r_1$ genes are 1, it indicates that the genes produce signaling molecules. If the states are 0, the genes do not synthesize signaling molecules.
- $r_2$ receptors: The states of $r_2$ receptors are determined by the average concentration of the signaling molecules within the neighborhood radius ($R$). Figure 5 shows an example calculating the average concentration of the signaling molecules from the neighboring cells to determine the state of receptor *gene 2* of cell $i$. Based on a certain threshold ($\tau_{th}$), the state of the receptor *gene 2* is updated. If the average concentration value is bigger than $\tau_{th}$, the state becomes 1. Otherwise, it becomes 0.

The following processes are taken to change the gene expression values which is caused by cell-cell interaction, and to determine a cell fate from the change of the gene values:

(1) Look into whether there are neighboring cells within the neighborhood radius ($R$) or not. In the case that there are neighbors, paracrine signaling is used. Otherwise, autocrine signaling is used.
(2) Determine the states of receptors ($r_2$) through the average concentrations of signaling molecules and the threshold $\tau_{th}$.
(3) Activate or inhibit genes that have the receptors as input nodes in a GRN based on the states of the receptors ($r_2$). If the states of receptors are 1, the states of genes become ON (1, activated). Otherwise, the states become OFF (0, inhibited).
(4) Check the attractor which the updated gene states finally converge to.
(5) Express the cell fate which is assigned to the attractor for cellular behaviors.
(6) Assign the states of the attractor as gene expression values for the next time step.

### 3.3 Cellular Movements

Our mechanism for cellular movements is based on Doursat's approach [11]. We determine cells' positions in each time step through Spring-mass-damper (SMD) kinetics. Specifically, all the cells have their positions in a Cartesian coordinate system. We assume that cells within the neighborhood radius ($R$) are connected by a spring with spring constant $k$ and equilibrium length $l$, and a damper with damping coefficient $c$ between each other. When cell A's position is $P_A = (x_A, y_A)$ and cell B's position is $P_B = (x_B, y_B)$, the equation for cellular movements is as follows:

$$m\ddot{P}_{AB} = -k\left(1 - \frac{l}{\|P_{AB}\|}\right)P_{AB} - c\dot{P}_{AB}$$

where

$$P_{AB} = P_B - P_A = (\delta cos\theta, \delta sin\theta),$$

$$\delta = \|P_{AB}\|, \theta = arctan\left(\frac{y_B - y_A}{x_B - x_A}\right),$$

$$\|P_{AB}\| = \|P_B - P_A\| = \sqrt{(x_B - x_A)^2 + (y_B - y_A)^2}$$

Because we neglect the effect of inertia, we replace $m\ddot{P}_{AB}$ with zero. Then, we finally obtain the following position update equation at each time step $\Delta t = 1$:

$$\Delta P_A = -\Delta P_B = \frac{\Delta P_{AB}}{2} = \frac{-k}{2c}(1 - \frac{l}{\|P_{AB}\|})P_{AB}$$

We can obtain different shapes of spatial patterns by the above position updating rule allowing physical interactions such as pushing or adhesion.

To acquire much more diverse spatial patterns, we introduce the dependence of parameters $k$, $l$, and $c$ on cell fates and perturbations to the cell position $(x, y)$ values. In the case of parameters $k$, $l$, and $c$, we determine the values depending on six types of cell fates possible between two cells: [*proli-proli*], [*proli-diff*], [*proli-qui*], [*diff-qui*], [*diff-diff*], and [*qui-qui*], where *proli* is proliferation, *diff* is differentiation, and *qui* is quiescence (Figure 6). Here apoptosis is excluded because cells due to apoptosis disappear in the space. In each simulation run, the parameter values of $k$, $l$, and $c$ are randomly chosen in certain ranges in Table1. For the perturbations, small perturbation values are added to the updated cell positions.

when the dependence of $k$, $l$, and $c$ on cell fates and perturbations to the cell positions are introduced, the final position of cell A whose neighbor is cell B is as follows:

$$P_A(t + 1) = P_A(t) + (\Delta P_A)_{[\alpha-\beta]} + \omega_{[\alpha-\beta]}$$

where α is cell A's cell fate, β is cell B's cell fate, and ω is the perturbation value to the updated coordinate of cell A.

## 4. Experiments

We performed 10,000 independent simulation runs for $K=$ 1, 2, 3, 4. Specifications of parameters for the simulations were the following:

- Space: the cells were positioned in 2D continuous 700 × 700 (in arbitrary unit) square area.
- Limitation of cell population: the cell population growth was limited up to 200 for reasonable computational loads in each run.
- Simulation termination condition: the simulations were terminated when the time step ($t$) was 1,000 or there was no cell in the space due to apoptosis.
- Parameter values of model: the values of parameters concerning GRN, Cell-Cell Interaction, and Cellular Movements follow Table 1.

### 4.1 Perturbations to GRNs and Robust & Evolvable GRNs

As mentioned in Introduction (Section 1), we have presented a new GRN-based morphogenetic model and revealed the role of the criticality of GRNs in morphogenesis [21, 22]. To make our finding more solid from an evolutionary viewpoint, we included genetic perturbations occurring in real living organisms in our morphogenetic model. Specifically, we perturbed the GRN of a seed cell by adding, deleting, or switching one regulatory link between a pair of genes. Based on the studies that biological systems have robustness and evolvability for evolution [8, 23, 28, 31, 40, 42], we focused on analyzing robust and evolvable GRNs among the perturbed GRNs. In our model, if the GRN conserves its existing attractors and create new attractors simultaneously against the perturbations, we considered the GRN as a robust and evolvable GRN (Figure 7) [2].

### 4.2 Morphological Measures for Pattern Analysis

We used 12 morphological measures (i - xii) to analyze morphogenetic patterns at the final time step based on our previous approaches [34, 35]. Among those measures, vi - xii are measures regarding network topology. To apply them to our morphologies, we constructed a network from each morphogenetic pattern. Concretely, we connected each cell to other cells within the neighborhood radius ($R$) in the space. Figure 8 shows an example morphology and a network constructed using our network construction method from it.

i. **Number of cells (*numOfCells*)**

ii. **Average distance of cells from center of mass (*massDistance*)**

The center of mass ($\bar{x}$, $\bar{y}$) was calculated from the means of $x$ and $y$ coordinates of all the cells. For average distance of cells from the center of mass, Euclidean distances between each cell position and the center of mass were measured.

iii. **Average pairwise distance (*pairDistance*)**

From two randomly sampled cells' positions, Euclidean distances were measured. The average was calculated based on 10,000 sampled pairs.

iv. **Kullback-Leibler divergence between pairwise particle distance distributions of a morphogenetic pattern and a random pattern (*kld*)**

This measure is to detect nontrivial morphologies. It was calculated from the Kullback-Leibler (KL) divergence between pairwise particle distance distributions of a morphogenetic pattern (Figure 9 (a)) and a randomly distributed pattern (Figure 9 (b)) made of the same number of cells within the same spatial dimensions. Each pairwise particle distance distribution was obtained through 10,000 random sampling of a pair of coordinates of cells (Figure 9 (c)).

v. **Mutual information between cell fates of cells and their neighboring cells (*MI*)**

This measure is to examine information of a correlation between the cell fates of cells and their neighboring cells in a morphogenetic pattern. It was calculated on the basis of the frequencies of neighboring cell fates. Figure 10 shows an example calculating *MI* in a morphogenetic pattern. In the case that there was only one cell, the value was set to 0.

vi. **Average clustering coefficient (*avgCluster*)**

vii. **Link density (*linkDensity*)**

viii. **Number of connected components (*numConnComp*)**

If there was a single cell in the space, it was considered one connected component.

ix. **Average size of connected components (*meanSizeConnComp*)**

When there was only one connected component, the value was set to 0.

x. **Homogeneity of sizes of connected components (*homoSizeConnComp*)**

This measure was calculated as the value that one minus the normalized entropy in the distribution of sizes of connected components. When there was only one connected component, the value was set to 1.

xi. **Size of the largest connected components (*sizeLarConnComp*)**

xii. **Average size of connected components smaller than the largest one (*meanSizeSmaller*)**

For the measures from i to xii, in the case that there was no cell in the space, all the values were set to 0.

### 4.3 Measures to investigate the relationship between GRNs and Expressed Cell Fates

We measured basin entropy and cell fates entropy from the sizes of basins of attractions and cell fates distributed in a morphogenetic pattern in order to investigate the relationship between GRNs and expressed cell fates. Here basin and cell fates entropy were computed except for apoptosis to compare with *MI* which was calculated considering cells expressing proliferation, differentiation, and quiescence in a morphogenetic pattern.

i. **Basin entropy**

$$H_{basin} = -\sum_{\rho} P_\rho \cdot log_2 P_\rho$$

where $P_\rho$ is the size of the basin of the attractor $\rho$ (in which proliferation, differentiation, quiescence were assigned), divided by the sum of sizes of the basins except for the basin size of the attractor for apoptosis. Thus,

$$\sum_{\rho} P_{\rho} = 1$$

Originally, basin entropy was suggested by Krawitz as a measure of the complexity of information that a system is able to store in $NK$ Boolean networks [24]. We used it as a measure to examine the versatility of the three cell fates (proliferation, differentiation, quiescence).

ii. **Cell fates entropy**

$$H_{cell\ fates} = -\sum_{f} P_f \cdot log_2 P_f$$

where $P_f$ is the number of cells expressing a cell fate $f$ (proliferation, differentiation, quiescence), divided by the sum of numbers of cells except for cells expressing apoptosis in a morphogenetic pattern at the final time step. Hence,

$$\sum_{f} P_f = 1$$

In the case that there were no cells expressing proliferation (differentiation/ quiescence), its log value was set to 0. Also, when there was no cell in the space, the value was set to 0.

## 5. Results

Figure 11 (a) shows probabilities of producing robust and evolvable GRNs against perturbations for $K = 1, 2, 3, 4$ in 10,000 simulation runs. We found that robust and evolvable GRNs were generated with the highest probability at $K = 2$. To show visualized morphogenetic patterns for each $K$, we presented morphologies which were produced by robust and evolvable GRNs acquired from 500 simulation runs in Figure 11 (b). Like the probabilities of Figure 11 (a), the number of the morphogenetic patterns at $K = 2$ was largest.

We measured 12 morphological measures focusing on morphologies obtained from morphogenetic systems having evolutionarily meaningful robust and evolvable GRNs. Because the robust and evolvable GRNs were generated with different probabilities in 10,000 simulation runs, we applied 1,000 bootstrap sampling to the values of the 12 measures for comparison between groups ($K = 1 - 4$) with unequal sample sizes. Figure 12 indicates the comparison of means between groups for the measures, where Kruskal-Wallis and Nemenyi (as post-hoc analysis) tests were performed to show statistically significant differences among the groups. Furthermore, we acquired a correlation matrix to investigate correlations between the 12 measures. In Figure 13, seeing the row of *numOfCells*, we found that six measures were highly correlated to *numOfCells*: *avgCluster*, *massDistance*, *meanSizeConnComp*, *meanSizeSmaller*, *pairDistance*, and *sizeLarConnComp*. These correlations were found in Figure 12 as well. The values of *numOfCells* at $K = 1, 2, 3$ were similar but fell sharply at $K = 4$. This trend was also shown in the six measures highly correlated to *numOfCells*. Meanwhile, *kld, MI, homoSizeConnComp*, *linkDensity,* and *numConnComp* showed different trends. In Figure 12 (c), *kld* was highest at $K = 2$, which means that nontrivial morphogenetic patterns were generated most frequently when the GRNs were critical under the genetic perturbations. This result demonstrates that the role of criticality of GRNs maintains even in the presence of evolutionary perturbations. In addition, from *MI, homoSizeConnComp*, *linkDensity,* and *numConnComp*, we could find out two interesting properties of the nontrivial morphologies at the criticality.

      Firstly, certain combinations of cell fates between neighboring cells occurred most frequently. In Figure 12 (l), *MI* was highest at $K = 2$. It indicates that the fate of a cell is strongly correlated to the fate of its neighboring cell in a morphogenetic pattern. To examine the relationship between GRNs and expressed cell fates, we measured basin entropy and cell fates entropy focusing on three cell fates except for apoptosis because a morphogenetic pattern was composed of cells expressing only proliferation, differentiation, and quiescence (Figure 14). Our original expectation was that if the basins of attraction for the three cell fates were most evenly distributed at $K = 2$, the expressions of different cell fates would be maximally balanced in a morphogenetic pattern. However, the cell fates entropy value was highest at $K = 1$ although basin entropy was highest at $K = 2$. Thus, we found that the distributions of cell fates in a morphology cannot be simply expected from the basin sizes of a GRN at a single cell level. It is one of emergent properties at a multicellular level obtained by the effect of criticality of GRNs during the process of cell-cell interaction. The spatial arrangements of cells at $K = 2$ where certain combinations of cell fates between neighbors occurred most frequently must be likewise understood as a multicellular emergent property.

      Secondly, the nontrivial morphologies emerged among topologically homogeneous cell clusters. In (b), (d), (h) of Figure 12, $K = 1$ relatively had higher *homoSizeConnComp*, lower *linkDensity*, lower *numConnComp* values on average when compared to the corresponding measures at $K = 2, 3$, and 4. Here we simply expressed it as (high, low, low). Similarly, in the same order, $K = 2$ had (high, high, high), $K = 3$ had (low, high, high), and $K = 4$ had (low, low, low). Based on the values of the measures, we found overall topological properties for

$K$ = 1 - 4. Specifically, the morphologies at $K$ = 1 consisted of homogeneous big size connected components which had a shape like a long chain. The morphologies at $K$ = 2 were composed of homogeneous size connected components where cells were interconnected. The morphogenetic patterns at $K$ = 3 were comprised of heterogeneous size connected components where cells were interconnected. The morphogenetic patterns at $K$ = 4 were made up of heterogeneous small size connected components which had a shape like a short chain. Figure 15 summarizes the topological properties for $K$ = 1 - 4 schematically.

In our morphogenetic process, there is a feedback relationship (Figure 16). Interactions with neighboring cells determine a cell fate. Depending on the cell fates, the parameters of SMD kinetics are applied. The cells are positioned by SMD kinetics. The positions of cells influence the number of neighboring cells. In the feedback relationship, we found that the nontrivial morphologies were produced most frequently when the GRNs were critical under the genetic perturbations. Besides, the nontrivial morphologies at the criticality had the most frequent certain combinations of cell fates between neighbors, and were composed of topologically homogeneous cell clusters. Because the parameter values of SMD kinetics determining cells' positions depended on the cell fates, the more certain combinations of cell fates between neighbors were, the more likely to be applied the same SMD parameter values were among the cells. Thus, the most frequent certain combinations of cell fates between neighbors would help the equal size connected components where cells were interconnected naturally emerge. Such spatial arrangements of cells due to the criticality of GRNs are not predictable properties at a single cell level. That is why studies on the role of criticality of GRNs in the context of multicellular settings are needed.

## 6. Conclusions

In this study, we introduced genetic perturbations to our morphogenetic systems having Kuffman's RBNs as GRNs and using SMD kinetics for cellular movements. We looked into whether or not the existing role of criticality of GRNs could be maintained even in the presence of the evolutionary perturbations and what the resulting morphogenesis were like. Our morphogenetic model was categorized into three according to the properties of GRNs. The dynamics of GRNs were varied from ordered ($K$ = 1), through critical ($K$ = 2), to chaotic ($K$ = 3, 4) regimes by node in-degree ($K$). We found that nontrivial morphologies were generated most frequently when the GRNs were critical under the genetic perturbations in common with the result in the morphogenetic systems without the evolutionary perturbations. Moreover, we found that the nontrivial morphologies at the criticality emerged among topologically homogeneous cell clusters due to the spatial arrangements in which certain combinations of cell fates between neighboring cells occurred most frequently. Based on these findings, we conclude that the criticality of GRNs facilitates the formation of nontrivial morphologies by adjusting the spatial arrangements of cells in GRN-based morphogenetic systems under the genetic perturbations.

Our findings correspond to an epigenetic viewpoint. Researchers in epigenesis have believed that heterogeneous and complex features emerge from homogeneous and less complex components through the interactions among them [27, 33]. In our model, we showed that the nontrivial morphologies at the criticality were produced most frequently among topologically homogeneous cell clusters. Thus, the result not only strongly supports the theory of epigenesis in developmental biology but also implies that highly structured tissues or organs in morphogenesis of multicellular organisms might stem from the criticality of GRNs.

For further study, we will track the growing processes from the seed cell to the cell aggregation to fully account for the spatial and temporal distribution of cells by the criticality of GRNs in morphogenesis. Furthermore, we plan to improve our morphogenetic model more biologically using real biological Boolean networks and the mechanisms for cellular movements such as a nutrient or oxygen gradient. In addition, we will explore its applicability as a framework which can quantitatively analyze morphological properties in the area of systems/computational biology.

## Acknowledgments

This material is based upon work supported by the US National Science Foundation under Grant No. 1319152.

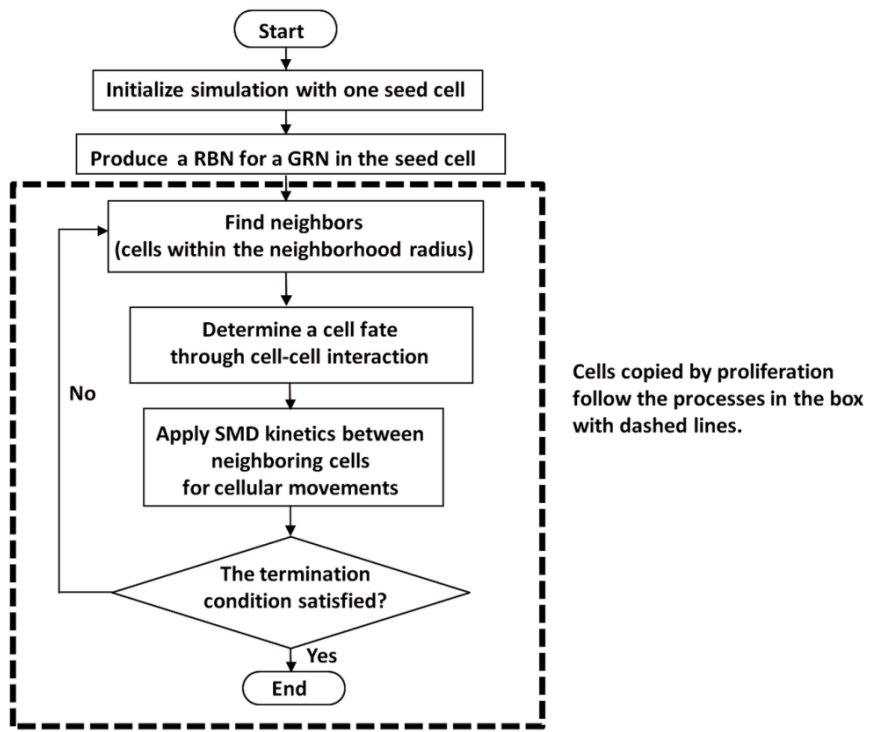

Figure 1. Algorithm for our GRN-based morphogenetic model

(a)

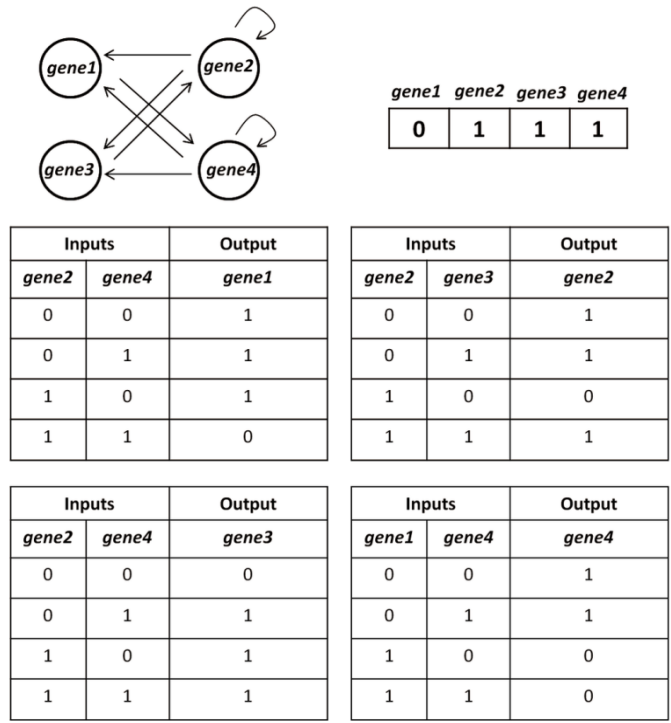

(b)

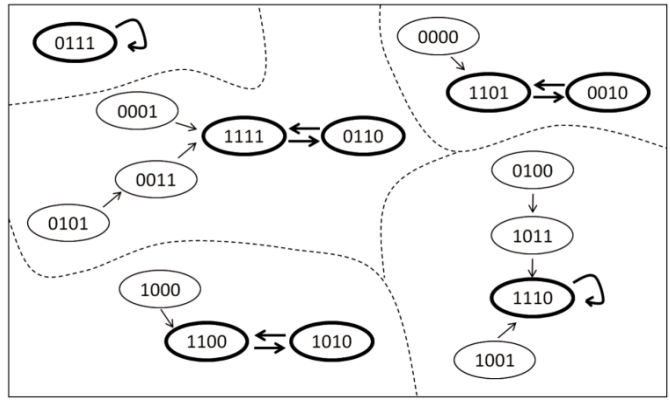

Figure 2. Schematic diagrams for an example GRN and its state space. (a) A RBN with $N$ = 4, $K$ = 2, and Boolean functions randomly assigned to each node. (b) State space of the RBN. The state space consists of $2^4 = 16$ configurations and transitions among them. The configurations with bold lines are attractors. Dashed lines draw boundaries for each basin of attraction.

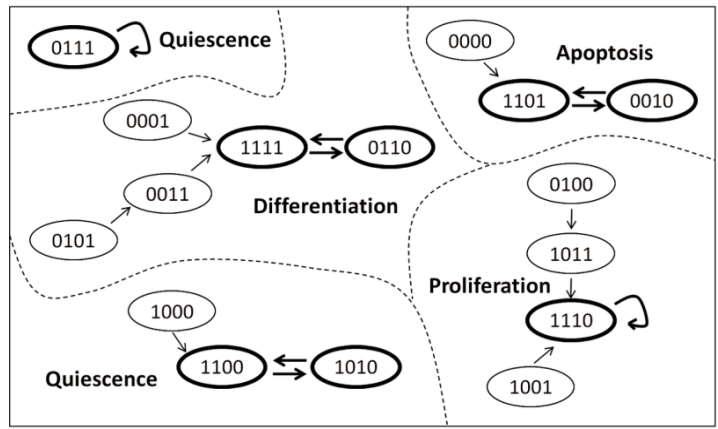

**Figure 3.** An example State space of a GRN where four cell fates are randomly assigned. (In actual simulations, 16 nodes were used. Thus, $2^{16} = 65{,}536$ configurations exist in the state space.)

(a)

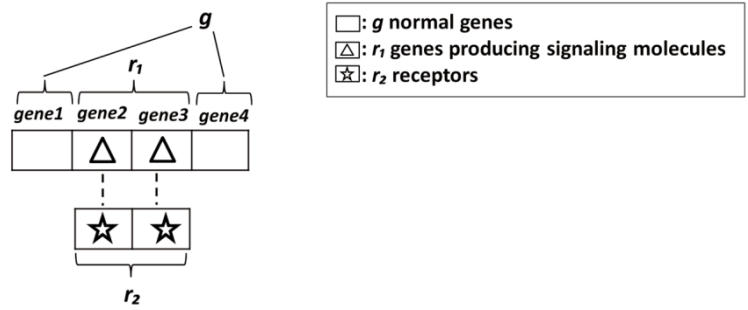

(b)

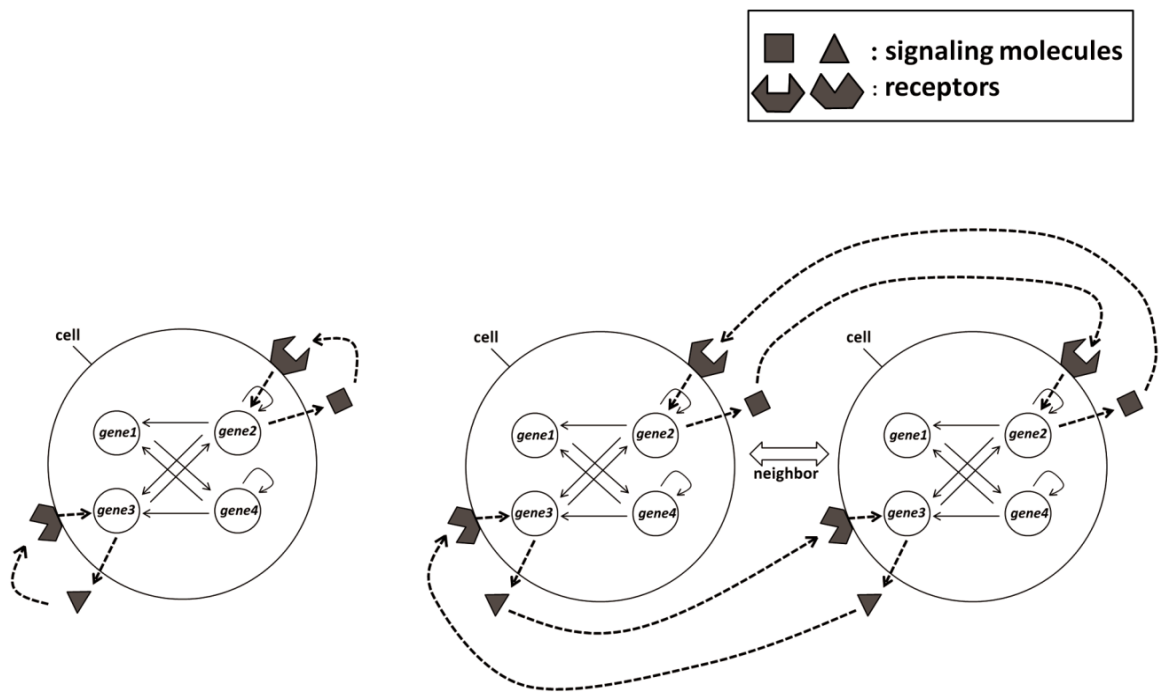

Figure 4. Cell signaling for cell-cell interaction. Schematic diagrams are examples illustrating the concept of cell signaling. (a) Assignment of genes in a GRN for cell signaling. (b) Two signal transduction mechanisms: autocrine (left) and paracrine (right).

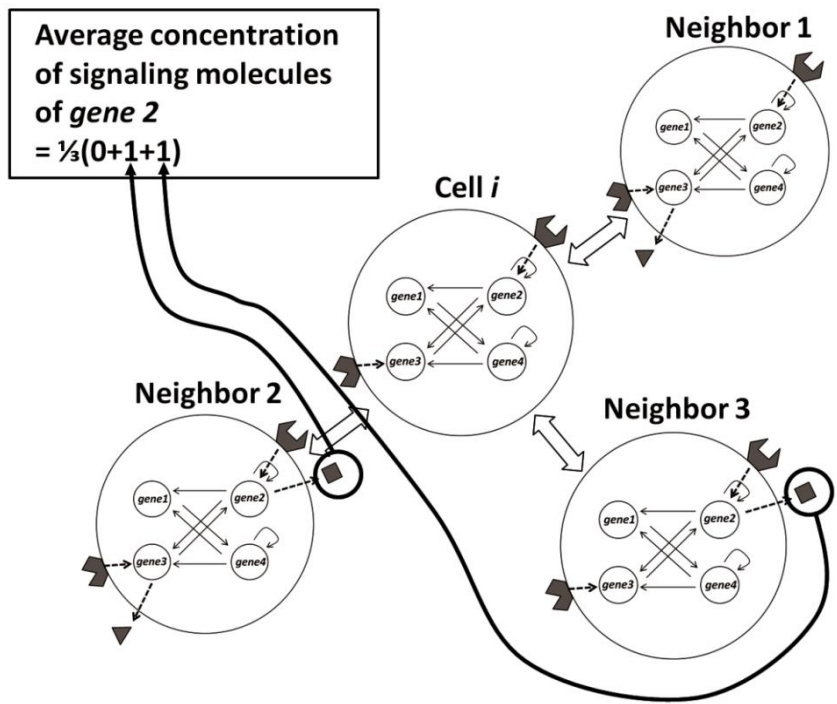

**Figure 5.** An example showing how to calculate the average concentration of the signaling molecules neighboring cells produce.

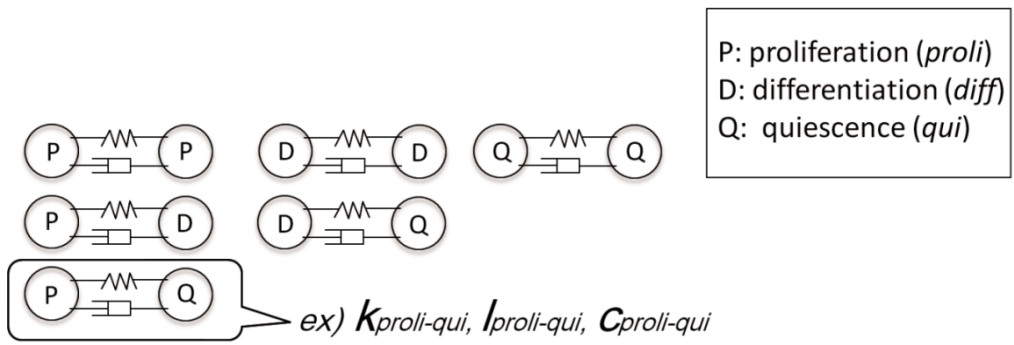

**Figure 6. Six types of cell fates possible between two cells.**

**Table 1. Parameters of model and their values**

| Model | Parameter | Value |
|---|---|---|
| GRN | Number of nodes ($N$) | 16 |
| | Number of in-degree per node ($K$) | 1, 2, 3, 4 |
| Cell-Cell Interaction | Neighborhood radius ($R$) | 30 |
| | Number of special genes ($r$) | 2 |
| | Threshold of signaling molecules ($\tau_{th}$) | 0.5 |
| Cellular Movements | Spring constant ($k$) | $\mathbb{R} \in$ unif (0, 1) |
| | Spring equilibrium length ($l$) | $\mathbb{R} \in$ unif (0, 100) |
| | Damper coefficient ($c$) | $\mathbb{R} \in$ unif (0, 200) |

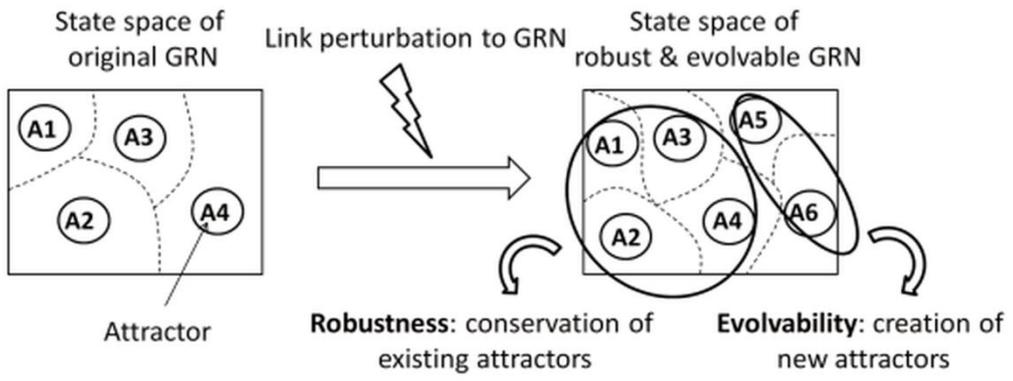

**Figure 7. Schematic diagrams illustrating the concept of a robust and evolvable GRN.**

(a) 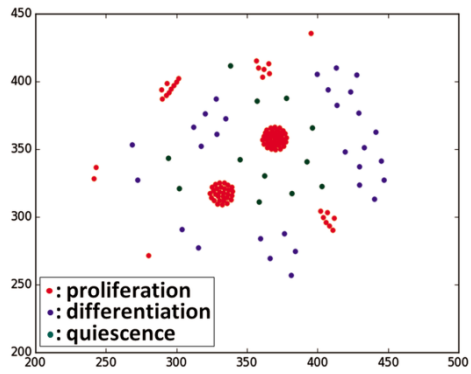 (b) 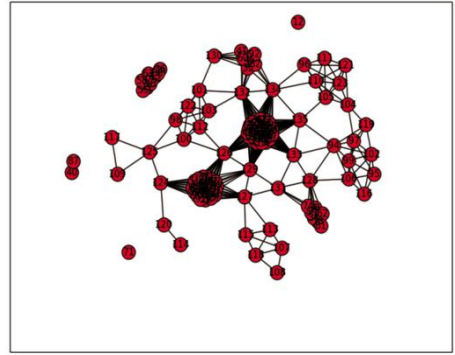

**Figure 8. Network construction for the analysis of morphologies. (a) Snapshot of a morphogenetic pattern. (b) Network constructed using our network construction method from (a).**

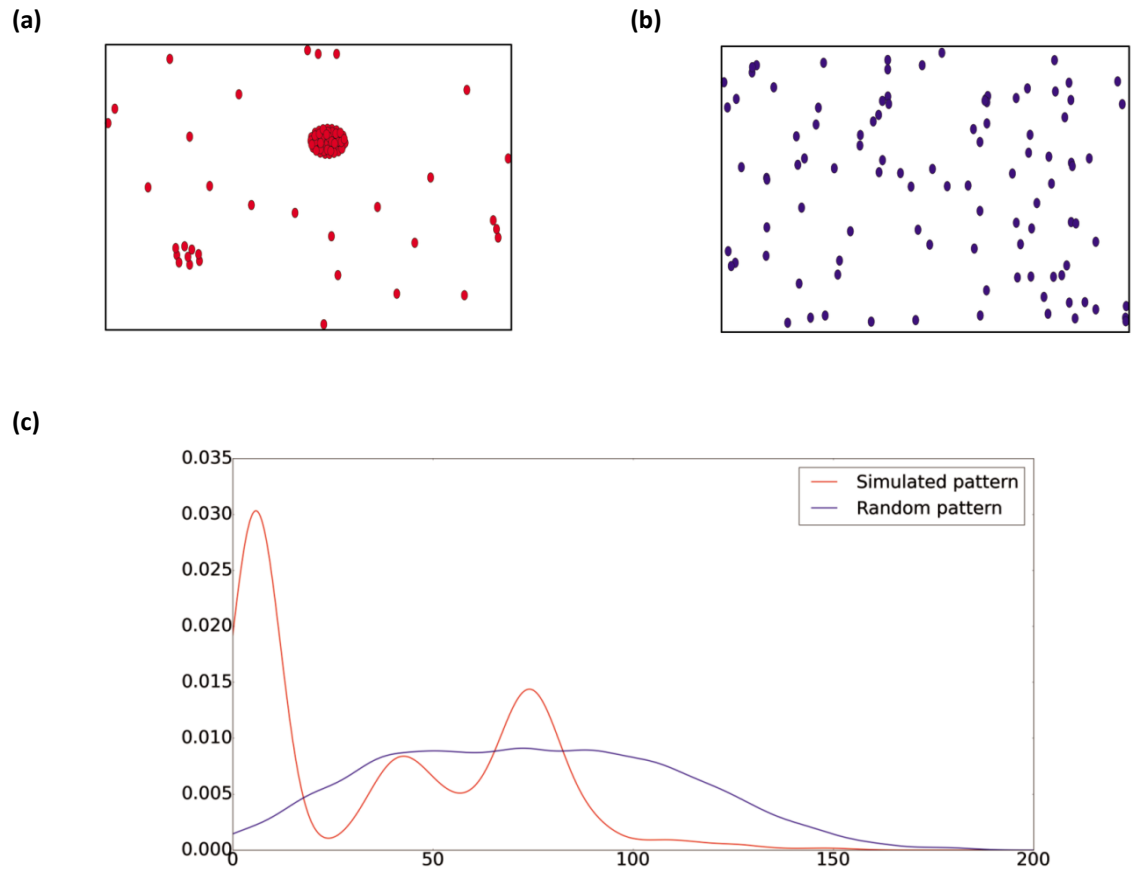

**Figure 9.** Nontrivial morphology detection using KL divergence. (a) A morphogenetic pattern acquired from a simulation. (b) A random pattern obtained from a uniform distribution. (c) Pairwise particle distance distributions of a simulated pattern and a random pattern. The curves are estimated by Gaussian kernel density estimation.

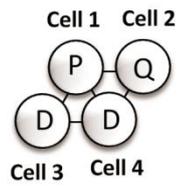

```
                Cell 1  Cell 2  Cell 3  Cell 4
X = [ P,  P,  P, Q, Q, D, D, D, D, D ]
Y = [ Q, D, D, P, D, P, D, P, Q, D ]

L = 10 (length of information)
MI = (H(X) + H(Y) - H(XY)) / (log L)
```

P: proliferation
D: differentiation
Q: quiescence

Figure 10. An example showing how to calculate mutual information (*MI*) between cell fates of cells and their neighboring cells. The value of computed mutual information was divided by log L for the purpose of normalization.

(a)

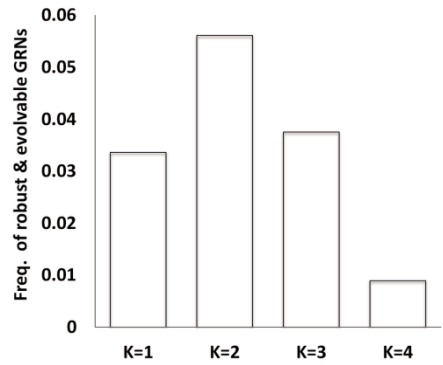

(b)

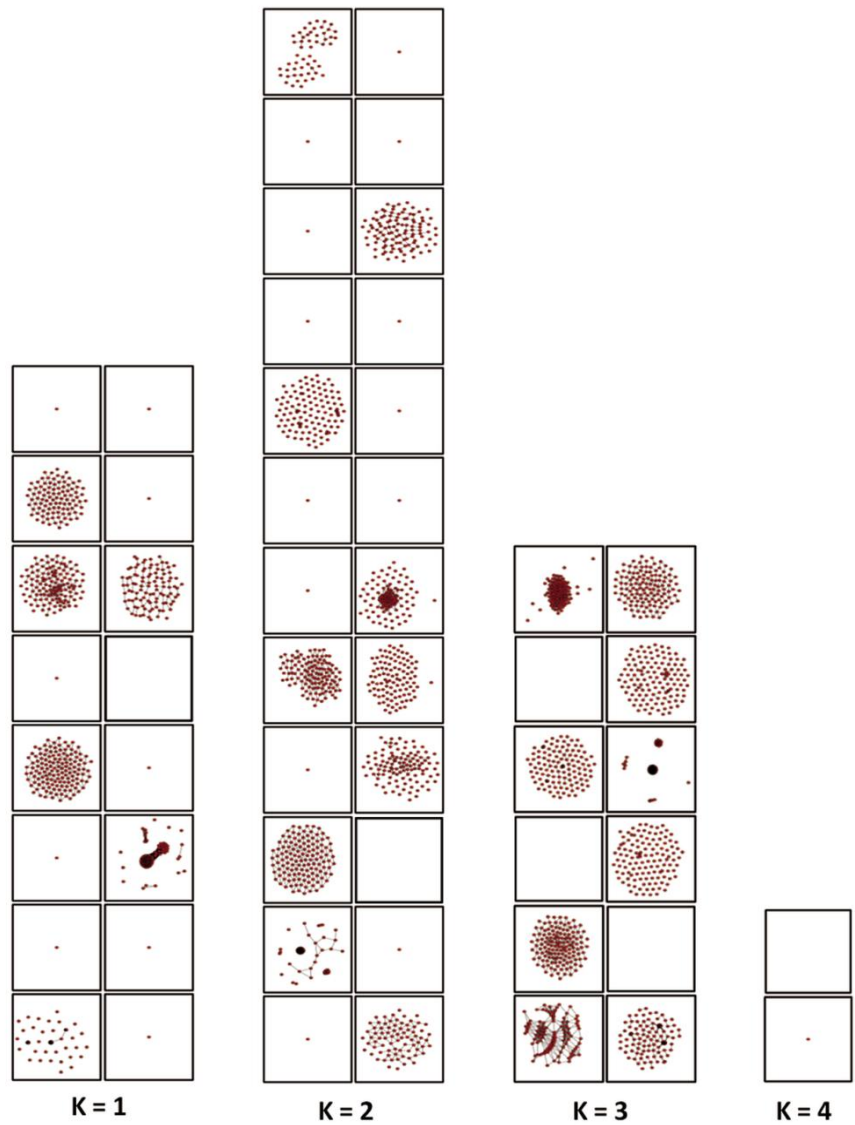

**Figure 11. (a) Probabilities of generating robust and evolvable GRNs for $K$ = 1, 2, 3, 4 in 10,000 simulation runs. (b) Different morphogenetic patterns obtained from robust and evolvable GRNs for $K$ = 1, 2, 3, 4. The numbers of the patterns were counted from robust and evolvable GRNs produced in 500 simulation runs.**

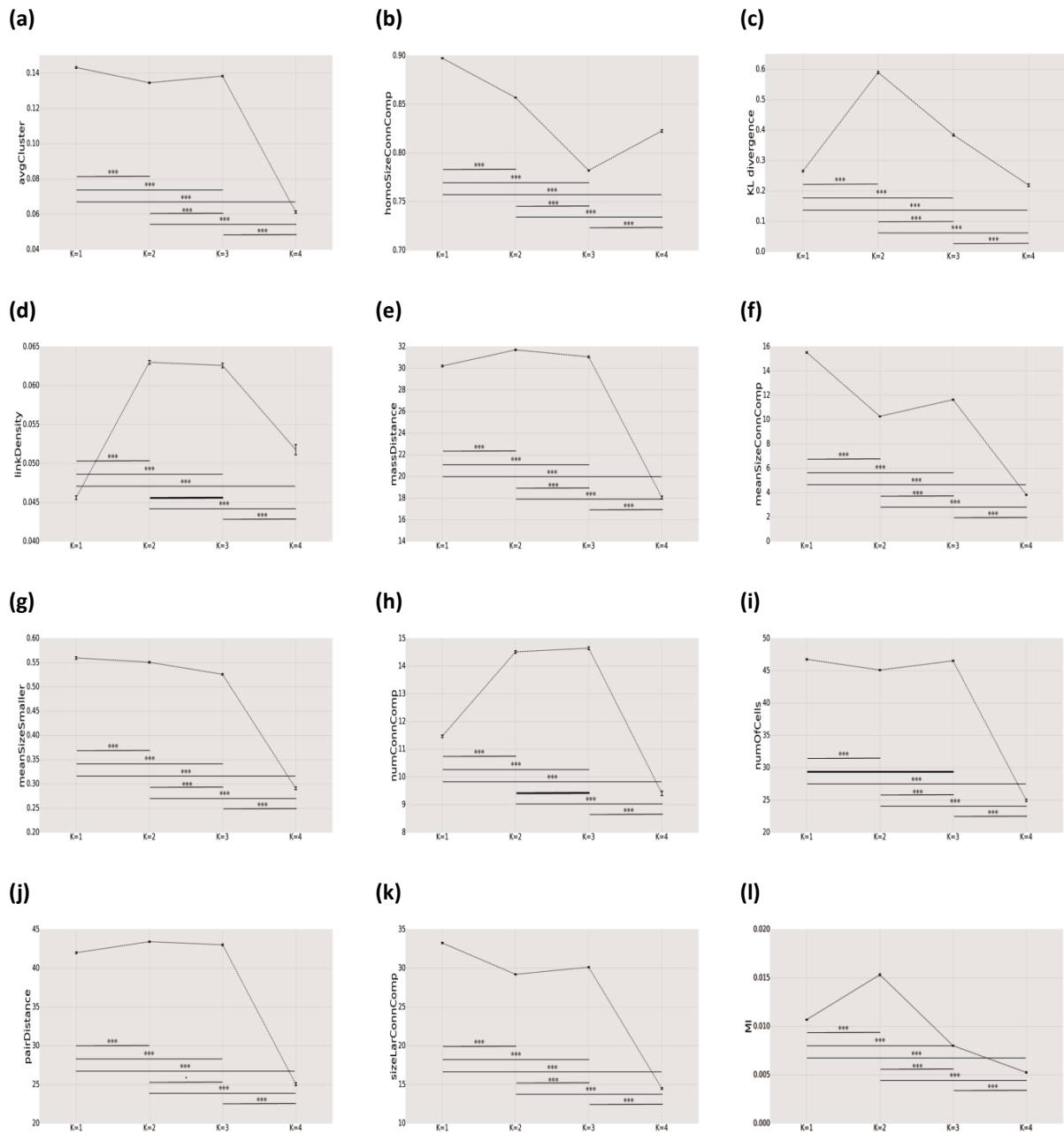

Figure 12. Comparison of means between groups ($K$ = 1, 2, 3, 4) for 12 morphological measures (Kruskal-Wallis test: $p < 2.2e\text{-}16$, Nemenyi test (post-hoc): ' ': $p < 1.0$, '***': $p < 0.001$). In the case that there is no difference between two groups, a bold line without an asterisk is presented in the plot. (a) Average clustering coefficient. (b) Homogeneity of sizes of connected components. (c) KL divergence between pairwise particle distance distributions of morphogenetic pattern and a random pattern. (d) Link density. (e) Average distance of cells from center of mass. (f) Average size of connected components. (g) Average size of connected components smaller than the largest one. (h) Number of connected components. (i) Number of cells. (j) Average pairwise distance. (k) Size of the largest connected component. (l) Mutual information between cell fates of cells and their neighboring cells.

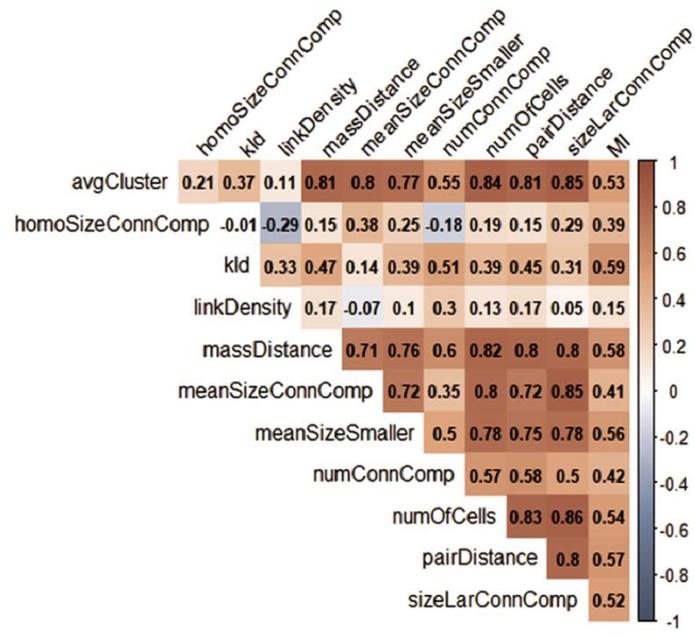

Figure 13. Correlation matrix for 12 morphological measures.

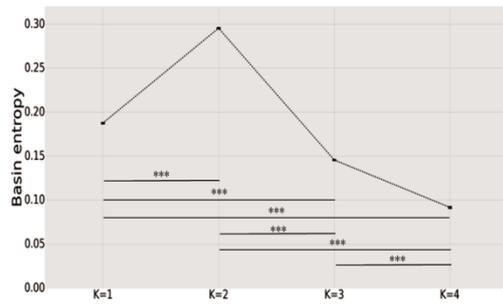 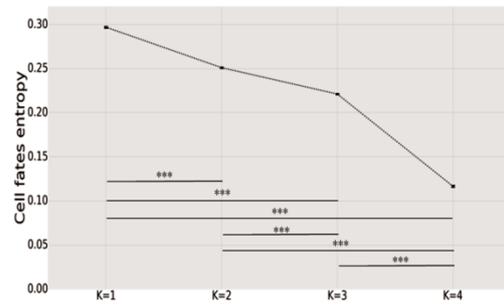

**Figure 14.** Comparison of means between groups for basin and cell fates entropy computed from three cell fates (proliferation, differentiation, quiescence). (a) Average basin entropy for K = 1, 2, 3, 4. (b) Average state entropy of cell fates performed in a simulation at the final time step for $K$ = 1, 2, 3, 4.

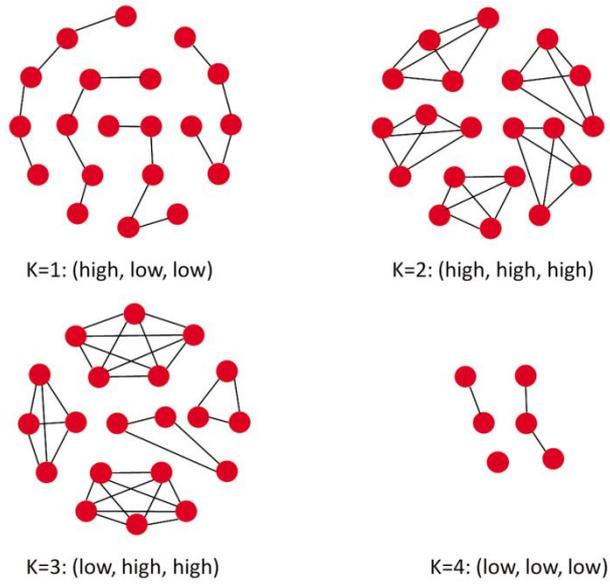

**Figure 15.** Topological properties of morphogenetic patterns for $K$ = 1, 2, 3, 4. "low" and "high" mean the relative values against $K$ in the order of (b) *homoSizeConnComp*, (d) *linkDensity*, (h) *numConnComp* in Figure 12.

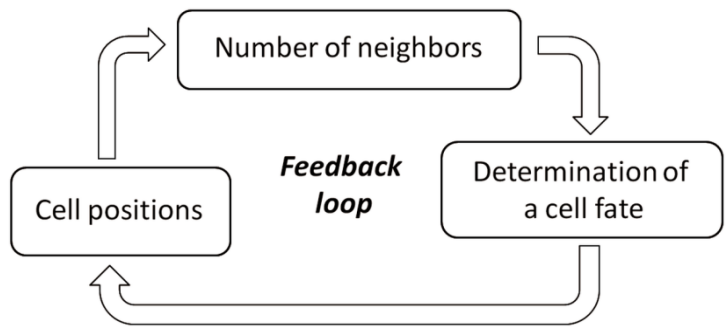

**Figure 16. Feedback relationship in our morphogenetic process.**